\documentclass[12pt]{article}

\usepackage{bbm,latexsym}

\newcommand{\be}{\begin{equation}}
\newcommand{\ee}{\end{equation}}
\newcommand{\ba}{\begin{eqnarray}}
\newcommand{\ea}{\end{eqnarray}}
\newcommand{\no}{\nonumber\\}

\newcommand{\mnu}{\mathcal{M}_\nu}
\newcommand{\deltasol}{\Delta m^2_\odot}
\newcommand{\deltaatm}{\Delta m^2_\mathrm{atm}}
\newcommand{\thetasol}{\theta_\odot}
\newcommand{\thetaatm}{\theta_\mathrm{atm}}

\begin{document}

\title{\normalsize \hfill UWThPh-2005-15 \\[1cm]
\LARGE
A model realizing the Harrison--Perkins--Scott \\
lepton mixing matrix}
\author{Walter Grimus\thanks{E-mail: walter.grimus@univie.ac.at} \\
\setcounter{footnote}{6}
\small Institut f\"ur Theoretische Physik, Universit\"at Wien \\
\small Boltzmanngasse 5, A--1090 Wien, Austria \\*[3.6mm]
Lu\'\i s Lavoura\thanks{E-mail: balio@cftp.ist.utl.pt} \\
\small Universidade T\'ecnica de Lisboa
and Centro de F\'\i sica Te\'orica de Part\'\i culas \\
\small Instituto Superior T\'ecnico,
1049-001 Lisboa, Portugal \\*[4.6mm]}

\date{15 December 2005}

\maketitle

\begin{abstract}
We present a supersymmetric model
in which the lepton mixing matrix $U$ obeys,
at the seesaw scale,
the Harrison--Perkins--Scott \textit{Ansatz}---vanishing $U_{e3}$,
maximal atmospheric neutrino mixing,
and $\sin^2{\theta_\odot} = 1/3$
($\theta_\odot$ is the solar mixing angle).
The model features a permutation symmetry $S_3$
among the three lepton multiplets of each type---left-handed doublets,
right-handed charged leptons,
and right-handed neutrinos---and among three Higgs doublets
and three zero-hypercharge scalar singlets;
a fourth right-handed neutrino,
a fourth Higgs doublet,
and a fourth scalar singlet are invariant under $S_3$.
In addition,
the model has seven $\mathbbm{Z}_2$ symmetries,
out of which six do not commute with $S_3$.
Supersymmetry is needed in order to eliminate
some quartic terms from the scalar potential,
quartic terms which would make impossible to obtain
the required vacuum expectation values
of the three Higgs doublets and three scalar singlets.
The Yukawa couplings to the charged leptons are flavour diagonal,
so that flavour-changing neutral Yukawa interactions
only arise at loop level.
\end{abstract}

\newpage

\section{Introduction} \label{intro}

After years of strenuous experimental efforts,
physicists are now in possession of quite a lot of good-accuracy data
on the neutrino mass-squared differences
and on the lepton mixing matrix.
The $3\sigma$ values derived in~\cite{tortola}
from all existing experimental neutrino oscillation data are
\ba
& & 7.2 \times 10^{-5}\, \mathrm{eV}^2
< \deltasol <
9.1 \times 10^{-5}\, \mathrm{eV}^2,
\\
& & 1.4 \times 10^{-3}\, \mathrm{eV}^2
< \deltaatm <
3.3 \times 10^{-3}\, \mathrm{eV}^2,
\\
& & 0.23 < \sin^2{\thetasol} < 0.38,
\label{sol} \\
& & 0.34 < \sin^2{\thetaatm} < 0.68,
\label{atm} \\
& & \left| U_{e3} \right|^2 < 0.047,
\label{ue3}
\ea
where $\deltasol \equiv m_2^2 - m_1^2$ is the solar mass-squared difference,
$\deltaatm \equiv \left| m_3^2 - m_1^2 \right|$
is the atmospheric mass-squared difference,
$\thetasol$ is the solar mixing angle,
$\thetaatm$ is the atmospheric mixing angle,
and $U_{e3}$ is one of the elements of the lepton mixing matrix $U$.
As seen in~(\ref{sol})--(\ref{ue3}),
salient features of $U$ are:
\begin{itemize}
\item The solar mixing angle is large but not maximal,
its sine-squared being possibly equal to $1/3$.
\item The atmospheric mixing angle is large and maybe maximal,
its sine-squared being possibly equal to $1/2$.
\item $U_{e3}$ is small,
it possibly vanishes.
\end{itemize}
The Harrison--Perkins--Scott (HPS) \textit{Ansatz}~\cite{tribi1,tribi2}
for the lepton mixing matrix
incorporates the three likely values given above:
$\sin^2{\thetasol} = 1/3$,
$\sin^2{\thetaatm} = 1/2$,
and $U_{e3} = 0$.
It states that $U = X V X^\prime$,
where $X$ and $X^\prime$ are diagonal unitary matrices and
\be
V = \left( V_1, \, V_2, \, V_3 \right) ,
\quad
V_1 = \frac{1}{\sqrt{6}} 
\left( \begin{array}{c} 2 \\ -1 \\ -1 \end{array} \right) ,
\quad
V_2 = \frac{1}{\sqrt{3}} 
\left( \begin{array}{c} 1 \\ 1 \\ 1 \end{array} \right) ,
\quad
V_3 = \frac{1}{\sqrt{2}} 
\left( \begin{array}{c} 0 \\ -1 \\ 1 \end{array} \right) .
\ee
Since $V$ is an orthogonal matrix,
the relations
\be
V_j^T V_k = \delta_{jk} \quad \mathrm{for} \quad j, k \in
\left\{ 1, 2, 3 \right\}
\ee
hold.

It is relatively easy to devise models which simultaneously produce
a maximal $\thetaatm$
and a vanishing $U_{e3}$---see for instance~\cite{soft,d4}.
It is much more difficult to reproduce
the third feature of the HPS mixing matrix,
namely $\sin^2{\thetasol} = 1/3$.
Some schemes or models have recently appeared
which realize HPS mixing~\cite{feruglio,ma,ross,babu,zee};
for a review see~\cite{altarelli}.
With the exception of~\cite{ross} which uses $SU(3)$,
those models rely on an internal symmetry $A_4$.
They use the following method,
first put forward in~\cite{tribi1}.
Defining
\be
\omega \equiv \exp{\left( i\, \frac{2 \pi}{3} \right)}
= \frac{- 1 + i \sqrt{3}}{2},
\ee
they try to get at
\be
U = U_\ell U_\nu = \mathrm{diag} \left( 1, \, \omega, \, \omega^2 \right)
\times V \times \mathrm{diag} \left( 1, \, 1, \, -i \right),
\ee
where
\be
U_\ell = \frac{1}{\sqrt{3}} \left( \begin{array}{ccc}
1 & 1 & 1 \\ 1 & \omega & \omega^2 \\ 1 & \omega^2 & \omega
\end{array} \right)
\ee
is the unitary matrix diagonalizing the charged-lepton mass matrix,
and
\be
U_\nu = \frac{1}{\sqrt{2}} \left( \begin{array}{ccc}
1 & 0 & -1 \\ 0 & \sqrt{2} & 0 \\ 1 & 0 & 1
\end{array} \right)
\ee
is the unitary matrix diagonalizing the neutrino mass matrix.

We,
on the contrary,
shall produce in this paper a model wherein,
in a convenient weak basis,
the charged-lepton mass matrix is already diagonal from the start,
while the neutrino mass matrix in that weak basis,
$\mnu$,
is diagonalized by the HPS matrix:
\be
V^T \mnu \, V =
\mathrm{diag} \left( m_1 e^{i \varphi_1}, \, m_2 e^{i \varphi_2}, \,
m_3 e^{i \varphi_3} \right).
\ee
Instead of $A_4$ as internal symmetry,
our model is based on an extension of $S_3$.

The basic idea behind our model is the following.
Suppose that one has a seesaw mechanism~\cite{seesaw,seesaw1}, 
with a neutrino Majorana mass matrix of the form
\be
\label{mdm}
\mathcal{M}_{D+M} = \left( \begin{array}{cc}
0 & M_D^T \\ M_D & M_R
\end{array} \right)
\ee
producing an effective light-neutrino mass matrix
\be
\mnu = - M_D^T M_R^{-1} M_D.
\ee
Suppose that the neutrino Dirac mass matrix $M_D$
and the right-handed-neutrino Majorana mass matrix $M_R$
are of the form
\be
\label{massmatrices}
M_D = a \mathbbm{1}, \quad
M_R = \mu_0 \mathbbm{1} + \mu_1 A + \mu_2 B,
\ee
with complex parameters $a$ and $\mu_{0,1,2}$,
and
\be
\label{AB}
A = \left( \begin{array}{ccc}
0 & 1 & 1 \\ 1 & 0 & 1 \\ 1 & 1 & 0
\end{array} \right),
\quad
B = \left( \begin{array}{ccc}
0 & 0 & 0 \\ 0 & 1 & -1 \\ 0 & -1 & 1
\end{array} \right).
\ee
(In~(\ref{massmatrices}) and throughout this paper,
the unit matrix $\mathbbm{1}$ is always $3 \times 3$.)
Equation~(\ref{massmatrices}) should be understood as holding
in the weak basis where the charged-lepton mass matrix is diagonal,
as already stated before.
Now,
$V$ diagonalizes $M_R$,
because $V_{1,2,3}$ are simultaneous eigenvectors of $A$ and $B$:
\ba
M_R V_1 &=& \left( \mu_0 - \mu_1 \right) V_1, \\
M_R V_2 &=& \left( \mu_0 + 2 \mu_1 \right) V_2, \\
M_R V_3 &=& \left( \mu_0 - \mu_1 + 2 \mu_2 \right) V_3, \\
V^T M_R \, V &=& \mathrm{diag} \left(
\mu_0 - \mu_1, \, \mu_0 + 2 \mu_1, \, \mu_0 - \mu_1 + 2 \mu_2 \right).
\ea
Since $M_D \propto \mathbbm{1}$ commutes with $V$,
one obtains that in this case
$U$ is equal to $V$ times a diagonal unitary matrix,
the light-neutrino masses being
\be
\label{mmm}
m_1 = \left| \frac{a^2}{\mu_0 - \mu_1} \right| , \quad
m_2 = \left| \frac{a^2}{\mu_0 + 2 \mu_1} \right| , \quad
m_3 = \left| \frac{a^2}{\mu_0 - \mu_1 + 2 \mu_2} \right| .
\ee
One thus reproduces HPS mixing.

In section~\ref{model} of this paper we present our model:
its field content,
internal symmetries,
and Lagrangian.
In section~\ref{mixing} we assume a definite form for the VEVs
of some scalar fields of our model and therefrom derive
that the lepton mixing matrix is of the HPS form. 
In section~\ref{scalarsector}
we show that the scalar potential of the model
is capable of producing VEVs of the desired form.
We present our conclusions in section~\ref{conclusions}.

\section{The model} \label{model}

Next we incorporate the idea at the end of the previous section
into a model. 
It turns out that the model has to be
supersymmetric;\footnote{We assume $R$-parity invariance,
just as in the minimal supersymmetric extension of the Standard Model.}
the argument for this will be given in section~\ref{scalarsector}.
However,
for convenience and transparency,
we shall use, 
wherever possible,
a non-supersymmetric notation.

\paragraph{Fermions:}
For $\alpha = e, \, \mu, \, \tau$,
three left-handed lepton doublets $D_{L\alpha}$,
three right-handed charged leptons $\alpha_R$,
and four right-handed neutrinos $\nu_{R\alpha}$ and $N_R$.

\paragraph{Scalars:}
For $\alpha = e, \, \mu, \, \tau$,
three Higgs doublets $\phi_\alpha$ with hypercharge $+1$;
furthermore,
one Higgs doublet $\phi_\nu$ with hypercharge $-1$;\footnote{In supersymmetry,
in order to cancel anomalies,
one needs~\cite{drees}
an equal number of Higgs doublets with hypercharges $+1$ and $-1$.
Hence we must add to our model two more Higgs doublets with hypercharge $-1$.
This addition is inconsequential,
though,
provided all Higgs doublets with hypercharge $-1$
behave under the internal symmetries in the way $\phi_\nu$ does.}
also,
four complex singlets of $SU(2)$ with zero hypercharge:
$\chi_\alpha$ and $S$.

\paragraph{Discrete symmetries:}
\begin{itemize}
\item One permutation group $S_3$,
effecting \emph{simultaneous} permutations of the $D_{L\alpha}$,
$\alpha_R$,
$\nu_{R\alpha}$,
$\phi_\alpha$,
and $\chi_\alpha$.
\item Three $\mathbbm{Z}_2$ symmetries
\be
\mathbf{z}_\alpha: \quad
D_{L\alpha} \to - D_{L\alpha}, \ 
\alpha_R \to - \alpha_R, \
\nu_{R\alpha} \to - \nu_{R\alpha}, \
\chi_\alpha \to -\chi_\alpha.
\ee
Notice that
$\phi_\alpha$ does \emph{not} change sign under $\mathbf{z}_\alpha$.
The symmetries $\mathbf{z}_\alpha$
may be interpreted as being discrete lepton numbers.
\item Three $\mathbbm{Z}_2$ symmetries
\be
\mathbf{z}^h_\alpha: \quad
\alpha_R \to -\alpha_R, \ \phi_\alpha \to -\phi_\alpha.
\ee
The $\mathbf{z}^h_\alpha$ are meant to ensure that
the $\phi_\beta$ with $\beta \neq \alpha$
have no Yukawa couplings to $\alpha_R$.

\item One $\mathbbm{Z}_2$ symmetry
\be\label{zchi}
\mathbf{z}_\chi: \quad 
N_R \to -N_R, \
\chi_e \to -\chi_e, \
\chi_\mu \to - \chi_\mu, \
\chi_\tau \to - \chi_\tau.
\ee
\end{itemize}
Notice that the scalar singlet $S$
is invariant under all these symmetries.
The need for this singlet $S$ will be explained
in section~\ref{scalarsector},
just as the need for the symmetry $\mathbf{z}_\chi$.

\paragraph{Yukawa Lagrangian:}
The multiplets and symmetries of the theory lead to
the Yukawa Lagrangian 
\ba
\mathcal{L}_Y &=&
- y\, \sum_\alpha \bar D_{L\alpha} \phi_\alpha \alpha_R
- y^\prime\, \sum_\alpha \bar D_{L\alpha} \phi_\nu \nu_{R\alpha} 
\no & &
+ \frac{y_\chi^\ast}{2}\, \sum_\alpha \chi_\alpha^\ast
\left( \nu_{R\alpha}^T C^{-1} N_R + N_R^T C^{-1} \nu_{R\alpha} \right) 
\no & &
+ \frac{y_S^\ast}{2} \sum_\alpha
S^\ast \nu_{R\alpha}^T C^{-1} \nu_{R\alpha}
+ \frac{y_N^\ast}{2}\, S^\ast N_R^T C^{-1} N_R
+ \mbox{H.c.}
\ea
We avoid writing down terms like
$S \sum_\alpha \nu_{R\alpha}^T C^{-1} \nu_{R\alpha}$
and $S\, N_R^T C^{-1} N_R$
since the model is meant to be supersymmetric,
hence $S$ and $S^\ast$ are not equivalent.
The absence from $\mathcal{L}_Y$ of terms like
$\sum_{\alpha,\beta} A_{\alpha\beta}
\bar D_{L\alpha} \phi_\nu \nu_{R\beta}$
and
$\sum_{\alpha,\beta} A_{\alpha\beta}
\chi_\alpha^\ast N_R^T C^{-1} \nu_{R\beta}$
is justified by the ``discrete lepton number''
symmetries $\mathbf{z}_\alpha$,
which are allowed to be broken only softly (see below).

\paragraph{Charged-lepton masses:}
The three different charged-lepton masses $m_\alpha$
are generated by the three different vacuum expectation values (VEVs)
$v_\alpha \equiv \langle 0 \left| \phi^0_\alpha \right| 0 \rangle$.
One has $m_\alpha = \left| y v_\alpha \right|$.
The charged-lepton mass matrix is diagonal as a consequence
of the symmetries $\mathbf{z}_\alpha$.

\paragraph{Symmetry breaking:}
\begin{itemize}
\item $S_3$ is broken softly,
by terms of dimension 2 \emph{but not by terms of dimension 3},
to its subgroup $S_2$,
the $\mu$--$\tau$ interchange symmetry.
This $S_2$ is broken only spontaneously.
\item The $\mathbf{z}_\alpha$ are broken softly
by terms both of dimension 3 and of dimension 2.
In particular,
they are broken by the bare Majorana masses
of the $\nu_{R\alpha}$,
see~(\ref{maj}) below.
This soft breaking by the Majorana masses
corresponds to the soft breaking of the family lepton numbers
in the models of~\cite{soft,soft2}.
\item The $\mathbf{z}^h_\alpha$ are broken softly
by terms both of dimension 3 and of dimension 2.
In particular,
they are broken by the term $\phi_\nu \sum_\alpha \phi_\alpha$
in the scalar potential,
see~(\ref{W1}) below.
\item $\mathbf{z}_\chi$ is broken \emph{only spontaneously}.
\end{itemize}

\paragraph{Mass Lagrangian:}
There are three Majorana mass terms for the right-handed neutrinos
allowed by the symmetries and their breaking;
the corresponding mass Lagrangian is
\be
\label{maj}
\mathcal{L}_M = 
\frac{\mu_0^\ast}{2}\, 
\sum_\alpha
\nu_{R\alpha}^T C^{-1} \nu_{R\alpha}
+ \frac{\mu_1^\ast}{2}\, \sum_{\alpha,\beta} A_{\alpha\beta}\,
\nu_{R\alpha}^T C^{-1} \nu_{R\beta}
+ \frac{m_N^\ast}{2}\, N_R^T C^{-1} N_R
+ \mbox{H.c.}
\ee
The matrix $A$ is given in~(\ref{AB}).
The second term breaks softly all three $\mathbf{z}_\alpha$.
Notice that we allow $S_3$ to be broken softly by terms of dimension 2,
but not by terms of dimension 3;
else the second term in~(\ref{maj}) would be more general.
Majorana mass terms of the type $N_R^T C^{-1} \nu_{R\alpha}$
are forbidden by $\mathbf{z}_\chi$,
which is not allowed to be broken in the Lagrangian.

In the supersymmetric theory,
(\ref{maj}) must be interpreted as part of the superpotential,
and gives rise to both dimension-3 and dimension-2 terms.

\paragraph{Superpotential:}
Let us
denote the superfield versions of $\chi_\alpha$,
$S$,
$\phi_\alpha$,
and $\phi_\nu$ by $\hat \chi_\alpha$,
$\hat S$,
$\hat\phi_\alpha$,
and $\hat\phi_\nu$, 
respectively.
Then the part of the superpotential relevant for the scalar potential is
\ba
\label{W1}
W &=&
\mu \hat \phi_\nu \sum_\alpha \hat \phi_\alpha
+ \mu^\prime \sum_\alpha \hat \chi_\alpha \hat \chi_\alpha
+ \mu^{\prime\prime} \sum_{\alpha, \beta}
A_{\alpha \beta} \hat \chi_\alpha \hat \chi_\beta
+ \mu^{\prime\prime\prime} \hat S \hat S
\\ & &
\label{W2}
+ \lambda \hat S \sum_\alpha \hat \chi_\alpha \hat \chi_\alpha
+ \lambda^\prime \hat S \hat S \hat S.
\ea
The symmetry $\mathbf{z}_\chi$
forbids odd powers of the fields $\hat \chi_\alpha$.
The symmetries $\mathbf{z}_\alpha^h$ forbid
a term $\hat S\, \hat\phi_\nu \sum_\alpha \hat \phi_\alpha$.
Equation~(\ref{W1}) gives rise to both
dimension-2 and dimension-3 terms;
(\ref{W2}) yields dimension-3 and dimension-4 terms.

\section{Lepton mixing} \label{mixing}

We denote the VEV of $\phi_\nu$ by $v$
and the VEVs of the $\chi_\alpha$ by $w_\alpha$.
We define  $a = {y^\prime}^\ast v^\ast$,  
and the $1 \times 3$ row matrix
\be
M = y_\chi \left( w_e, \, w_\mu, \, w_\tau \right).
\ee
Then,
the $7 \times 7$ neutrino Majorana mass matrix,
in the basis $\left[ \nu_{L\alpha}, \,
\left( \nu_{R\alpha} \right)^c, \,\left( N_R \right)^c \right]$,
is
\be
\label{D+M}
\mathcal{M}_{D+M} = \left(
\begin{array}{ccc}
0_{3 \times 3} & a \mathbbm{1} & 0_{3 \times 1} \\
a \mathbbm{1} & \bar \mu_0 \mathbbm{1} + \mu_1 A & M^T \\
0_{1 \times 3} & M & \bar m_N
\end{array} \right),
\ee
where $\bar \mu_0 \equiv \mu_0 + y_S s$
and $\bar m_N \equiv m_N + y_N s$,
$s$ being the VEV of $S$. 
The size of the zero matrices is indicated as a subscript.
Comparing~(\ref{D+M}) with~(\ref{mdm}) gives 
\be
\label{MRD}
M_R = \left( \begin{array}{cc}
\bar \mu_0 \mathbbm{1} + \mu_1 A & M^T \\ M & \bar m_N
\end{array} \right)
\quad \mbox{and} \quad
M_D = \left( \begin{array}{c} a \mathbbm{1} \\
0_{1 \times 3} \end{array} \right).
\ee

In order to obtain the HPS mixing matrix
we must now make the following \emph{crucial assumption}
about the VEVs of the $\chi_\alpha$:
\be
\label{VEVs}
w_e = 0, \quad w_\mu = - w_\tau \equiv w.
\ee
In the next section we shall discuss
the conditions under which these relations can be obtained.
With~(\ref{VEVs}) we may write 
\be
\label{M}
M = \left( 0, \, m, \, - m \right),
\quad \mathrm{where} \ m = y_\chi w.
\ee

What is \emph{the idea} behind this construction?
First suppose that in the $M_R$ of~(\ref{MRD}) the entry $\bar m_N$
is much larger than all other entries.
Then,
there is a seesaw mechanism
\emph{within the $4 \times 4$ matrix $M_R$ itself},
generating an effective $3 \times 3$ matrix 
$\bar \mu_0 \mathbbm{1} + \mu_1 A - M^T M \left/ \bar m_N \right.$,
which is exactly of the form proposed in~(\ref{massmatrices})
provided $M$ is of the form in~(\ref{M}).
Next let us admit that $\bar m_N$ is \emph{not} much larger
than the other entries of $M_R$;
we shall demonstrate in the rest of the present section
that \emph{HPS mixing is realized even in that general case}.

We define the vectors
\be
W_j = \left( \begin{array}{c} V_j \\ 0 \end{array} \right)
\ \mbox{for} \ j = 1,2,3, \quad 
W_4 = \left( \begin{array}{c} 0_{3 \times 1} \\ 1 \end{array} \right),
\ee
and the real orthogonal matrix
$W = \left( W_1, \, W_2, \, W_3, \, W_4 \right)$.
We compute
\be
W^T M_R \, W = \left( \begin{array}{cccc}
\bar \mu_0 - \mu_1 & 0 & 0 & 0 \\
0 & \bar \mu_0 + 2 \mu_1 & 0 & 0 \\
0 & 0 & \bar \mu_0 - \mu_1 & -\sqrt{2} m \\
0 & 0 & -\sqrt{2} m & \bar m_N
\end{array} \right).
\ee
We diagonalize the $2 \times 2$ submatrix of $W^T M_R \, W$ as
\be
K^T \left( \begin{array}{cc}
\bar \mu_0 - \mu_1 & -\sqrt{2} m \\ -\sqrt{2} m & \bar m_N
\end{array} \right) K = 
\left( \begin{array}{cc} \bar m_3 & 0 \\ 0 & \bar m_4 \end{array}
\right),
\ee
where $K$ is a $2 \times 2$ unitary matrix
and $\bar m_{3,4}$ are real and non-negative. 
It is convenient to denote the indices of $K$ by 3 and 4:
\be
K = \left( \begin{array}{cc}
K_{33} & K_{34} \\ K_{43} & K_{44}
\end{array} \right).
\ee
Then one has
\ba
M_R^{-1} &=&
\frac{1}{\bar \mu_0 - \mu_1}\, W_1 W_1^T + 
\frac{1}{\bar \mu_0 + 2 \mu_1}\, W_2 W_2^T
\no & &
+ \frac{1}{\bar m_3}
\left( K_{33} W_3 + K_{43} W_4 \right)
\left( K_{33} W_3^T + K_{43} W_4^T \right)
\no & &
+ \frac{1}{\bar m_4}
\left( K_{34} W_3 + K_{44} W_4 \right)
\left( K_{34} W_3^T + K_{44} W_4^T \right)
\ea
and
\be
\mnu =
- \frac{a^2}{\bar \mu_0 - \mu_1}\, V_1 V_1^T
- \frac{a^2}{\bar \mu_0 + 2 \mu_1}\, V_2 V_2^T 
- a^2 \left[ \frac{ \left( K_{33} \right)^2}{\bar m_3} +
\frac{\left( K_{34} \right)^2}{\bar m_4} \right] V_3 V_3^T.
\ee
This is just of the desired form.
We have obtained an (indirect) realization
of the idea in section 1.

\section{The scalar sector and the VEVs} \label{scalarsector}

Recall the features of the VEVs that our model requires:
\begin{itemize}
\item The VEVs $v_\alpha$ of the $\phi_\alpha$
must be all different in order that
the charged-lepton masses are all different.
\item The VEVs $w_\alpha$ of the $\chi_\alpha$
must obey~(\ref{VEVs}) in order to obtain
the matrix $B$ of~(\ref{AB}) in $M_R$.
\end{itemize}

Now suppose that our model was not supersymmetric.
Then,
in the scalar potential,
terms like
\be
\label{forbidden}
\begin{array}{c}
\left( \phi_e^\dagger \phi_e \right) \left| \chi_e \right|^2
+
\left( \phi_\mu^\dagger \phi_\mu \right) \left| \chi_\mu \right|^2
+
\left( \phi_\tau^\dagger \phi_\tau \right) \left| \chi_\tau \right|^2,
\\*[2mm]
\left( \phi_\mu^\dagger \phi_\mu + \phi_\tau^\dagger \phi_\tau \right)
\left| \chi_e \right|^2
+
\left( \phi_e^\dagger \phi_e + \phi_\tau^\dagger \phi_\tau \right)
\left| \chi_\mu \right|^2
+
\left( \phi_e^\dagger \phi_e + \phi_\mu^\dagger \phi_\mu \right)
\left| \chi_\tau \right|^2
\end{array}
\ee
would be unavoidable.
Once all the $v_\alpha$ are different,
terms like the ones in~(\ref{forbidden})
force the $w_\alpha$ to be all different too
(unless they all vanish, which is not what we want).
This means that terms of dimension 4 in the scalar potential
which contain simultaneously the Higgs doublets
\emph{and} the $\chi$ singlets must be forbidden.
The only way to do this seems to be supersymmetry (SUSY).
In SUSY the scalar potential has three contributions:
\begin{itemize}
\item the D terms from the gauge interactions,
\item the F terms from the superpotential $W$, and
\item soft SUSY-breaking terms,
which are of dimension less than 4.
\end{itemize}
The $\chi$ fields have no D terms
since they are gauge singlets with hypercharge zero.
In the superpotential $W$,
the symmetries of our model do not allow terms
with products of $\chi$ fields and Higgs doublets.
Therefore,
once our model has been supersymmetrized,
its scalar potential has no unwanted dimension-4 terms.

The symmetry $\mathbf{z}_\chi$ in our model
is necessary in order to forbid terms like
\be
\label{forbidden3}
\left( \phi_e^\dagger \phi_e \right) \chi_e
+
\left( \phi_\mu^\dagger \phi_\mu \right) \chi_\mu
+
\left( \phi_\tau^\dagger \phi_\tau \right) \chi_\tau,
\ee
which would also be dangerous.
Unfortunately, $\mathbf{z}_\chi$ also forbids
cubic $\chi_\alpha$ terms
in the superpotential $W$;
this absence has as a consequence the absence
of quartic $\chi_\alpha$ terms in scalar potential.
Hence,
we have to introduce the scalar field $S$
in order to allow for the first term in~(\ref{W2}),
which gives then rise to a quartic $\chi_\alpha$ term
in the scalar potential.

The scalar potential of our supersymmetric model
is a sum  $V = V_S + V_{\phi S} + V_{\chi S}$,
where $V_S$ is a functions of $S$ alone,
$V_{\phi S}$ contains the pure Higgs doublet terms and the
mixed $S$--Higgs doublet terms,
and analogously for $V_{\chi S}$.
The full separation of $V_{\phi S}$ from $V_{\chi S}$
allows us to achieve completely different forms
for the VEVs of the $\phi$ doublets and $\chi$ singlets.
This is a crucial feature of our model.

\subsection{The VEVs of the $\chi_\alpha$}

The terms in the superpotential containing the fields $\hat \chi_\alpha$,
\be
\label{term1}
\left( \mu^\prime + \lambda \hat S \right)
\left( \hat \chi_e \hat \chi_e
+ \hat \chi_\mu \hat \chi_\mu
+ \hat \chi_\tau \hat \chi_\tau \right)
+ 2 \mu^{\prime\prime} \left( \hat \chi_e \hat \chi_\mu
+ \hat \chi_e \hat \chi_\tau
+ \hat \chi_\mu \hat \chi_\tau \right),
\ee
lead in $V_{\chi S}$ to the following quadratic terms:
\be
\begin{array}{c}
\left| \chi_e \right|^2
+ \left| \chi_\mu \right|^2
+ \left| \chi_\tau \right|^2,
\\*[1mm]
\mathrm{Re} \left(
\chi_e^\ast \chi_\mu + \chi_e^\ast \chi_\tau + \chi_\mu^\ast \chi_\tau
\right).
\end{array}
\ee
These may be added to the soft-breaking terms,
which are more general anyway.
But~(\ref{term1}) also leads to a \emph{quartic} term
\be
\lambda^2 \left| \left( \chi_e \right)^2
+ \left( \chi_\mu \right)^2
+ \left( \chi_\tau \right)^2 \right|^2
\ee
which is crucial,
since without this term the potential for the $\chi$ fields
would be purely quadratic
and all those fields would then be forced to have vanishing VEVs.
This is the crucial role played by $\hat S$ in our model.

Let us now consider how $V_{\chi S}$ depends
on the $\chi$ fields.\footnote{The dependence on the singlet $S$
is irrelevant for our purposes,
since we are only interested in seeing
whether it is possible to obtain
the desired VEVs for the $\chi_\alpha$.}
The symmetry $\mathbf{z}_\chi$,
which is broken only spontaneously,
forces $V_{\chi S}$ to contain only quadratic and quartic terms
on the $\chi$ fields.
The symmetry $S_3$ is broken softly,
in the quadratic terms,
to the $\mu$--$\tau$ interchange symmetry.
Hence,
if one defines $F \left( w_e, \, w_\mu, \, w_\tau \right)
\equiv \left\langle 0 \left| V_{\chi S} \right| 0 \right\rangle$,
then one has
\ba
F &=&
p \left| w_e \right|^2
+ q \left( \left| w_\mu \right|^2 + \left| w_\tau \right|^2 \right)
+ 2\, r\, \mathrm{Re} \left( w_\mu^\ast w_\tau \right)
+ 2\, \mathrm{Re} \left[ t\, w_e^\ast \left( w_\mu + w_\tau \right) \right]
\no & &
+ 2\, \mathrm{Re} \left[
P\, w_e^2
+ Q \left( w_\mu^2 + w_\tau^2 \right)
+ R\, w_\mu w_\tau
+ T\, w_e \left( w_\mu + w_\tau \right) \right]
\no & &
+ \lambda^2 \left| w_e^2 + w_\mu^2 + w_\tau^2 \right|^2.
\ea
The coefficients $p$,
$q$,
and $r$ are real,
all other coefficients are in general complex.
The coefficients of the terms quadratic in the $w_\alpha$
all contain some contributions including the VEV $s$ of the field $S$,
but it is useless,
for our purposes,
to explicitly consider that dependence.

The extremum conditions are
\ba
& & 0 = \frac{\partial F}{\partial w_e^\ast} =
p\, w_e
+ t \left( w_\mu + w_\tau \right)
+ 2 P^\ast w_e^\ast
+ T^\ast \left( w_\mu^\ast + w_\tau^\ast \right)
+ 2 \lambda^2\, w_e^\ast \left( w_e^2 + w_\mu^2 + w_\tau^2
\right),
\no & & \label{1} \\
& & 0 = \frac{\partial F}{\partial w_\mu^\ast} =
q\, w_\mu
+ r\, w_\tau
+ t^\ast\, w_e
+ 2 Q^*\, w_\mu^\ast
+ R^\ast\, w_\tau^\ast
+ T^\ast\, w_e^\ast + 
2 \lambda^2\, w_\mu^\ast \left( w_e^2 + w_\mu^2 + w_\tau^2
\right),
\no & & \label{2} \\
& & 0 = \frac{\partial F}{\partial w_\tau^\ast} =
q\, w_\tau
+ r\, w_\mu
+ t^\ast\, w_e
+ 2 Q^\ast\, w_\tau^\ast
+ R^\ast\, w_\mu^\ast
+ T^\ast\, w_e^\ast
+ 2 \lambda^2\, w_\tau^\ast \left( w_e^2 + w_\mu^2 + w_\tau^2
\right).
\no & & \label{3}
\ea
The desired solution of these equations is given in~(\ref{VEVs}).
Remarkably,
that \textit{Ansatz} automatically solves~(\ref{1}),
while~(\ref{2}) and~(\ref{3}) \emph{both} lead to
\be
\label{w}
\left( q - r \right) w
+ \left( 2 Q^\ast - R^\ast \right) w^\ast
+ 4 \lambda^2 \left| w \right|^2 w = 0.
\ee
Writing $w = w_0 e^{i \phi}$ with $w_0 > 0$, 
the relevant solution of~(\ref{w}) is
\be
\label{w0}
w_0^2 = \frac{r - q - \epsilon \left| 2Q - R \right|}{4 \lambda^2},
\quad
e^{2 i \phi} = \epsilon\, e^{- i \arg \left( 2 Q - R \right)},
\quad
\ee
where $\epsilon = \pm 1$.

Now the question is whether the above solution
of the extremum conditions is a local minimum of $F$.
To check this,
write
\be
w_e = e^{i \phi} \delta_e, \quad 
w_\mu = w + e^{i \phi  } \delta_\mu, \quad
w_\tau = - w + e^{i \phi} \delta_\tau,
\ee
where the factors $e^{i \phi}$ have been inserted for convenience.
Collect all the terms of $F$
which are quadratic on the $\delta_\alpha$: 
\ba
F_2 \left( \delta \right) &=&
p \left| \delta_e \right|^2
+ q \left( \left| \delta_\mu \right|^2
+ \left| \delta_\tau \right|^2 \right)
+ 2\, r\, \mathrm{Re} \left( \delta_\mu^\ast \delta_\tau \right)
+ 2\, \mathrm{Re} \left[ t\, \delta_e^\ast
\left( \delta_\mu + \delta_\tau \right) \right]
\no & &
+ 2\, \mathrm{Re} \left[
P^\prime\, \delta_e^2
+ Q^\prime \left( \delta_\mu^2 + \delta_\tau^2 \right)
+ R^\prime\, \delta_\mu \delta_\tau
+ T^\prime\, \delta_e \left( \delta_\mu + \delta_\tau \right) \right]
\no & &
+ 4 \lambda^2 w_0^2 \left[ \mathrm{Re} \left(
\delta_e^2 + \delta_\mu^2 + \delta_\tau^2 \right)
+ \left| \delta_\mu - \delta_\tau \right|^2 \right],
\label{F2}
\ea
where we have defined $P^\prime \equiv P e^{2 i \phi}$,
etc.
One has to show that $F_2 \left( \delta \right)$
 is a positive quadratic form in the six variables 
$\mathrm{Re}\, \delta_\alpha$ and $\mathrm{Im}\, \delta_\alpha$,
$\alpha = e, \mu, \tau$.
One has first to insert into~(\ref{F2}) the $w_0^2$ of~(\ref{w0}).
Considering $\delta_e$ alone,
the parameter $p$ may be assumed positive
and as large as necessary for positive definiteness.
Then setting $\delta_e = 0$ and $R = 0$,
one has
\ba
F_2 \left( \delta_e = 0, \, R = 0 \right) &=&
\left( \begin{array}{cc}
\mathrm{Re}\, \delta_\mu, & \mathrm{Re}\, \delta_\tau
\end{array} \right)
\left( \begin{array}{cc}
2 r - q - 2 \epsilon \left| Q \right| & q + 2 \epsilon \left| Q \right| \\
q + 2 \epsilon \left| Q \right| & 2 r - q - 2 \epsilon \left| Q \right|
\end{array} \right) 
\left( \begin{array}{c}
\mathrm{Re}\, \delta_\mu \\ \mathrm{Re}\, \delta_\tau
\end{array} \right)
\no & & +
\left( \begin{array}{cc}
\mathrm{Im}\, \delta_\mu, & \mathrm{Im}\, \delta_\tau
\end{array} \right)
\left( \begin{array}{cc}
q - 2 \epsilon \left| Q \right| & q + 2 \epsilon \left| Q \right| \\
q + 2 \epsilon \left| Q \right| & q - 2 \epsilon \left| Q \right|
\end{array} \right) 
\left( \begin{array}{c}
\mathrm{Im}\, \delta_\mu \\ \mathrm{Im}\, \delta_\tau
\end{array} \right).
\ea
The positivity conditions for the second $2 \times 2$ matrix are 
$\epsilon = -1$ and $q > 0$.
The first $2 \times 2$ matrix produces the conditions
$r > 0$ and $r > q - 2 \left| Q \right|$.
Since the eigenvalues of a quadratic form
are continuous functions of its parameters,
switching on $R$ will leave the eigenvalues positive
provided $\left| R \right|$ is not too large.
We have thus proved that there is a region of parameters
of the potential for which there is a local minimum
of the form~(\ref{VEVs}).

\subsection{The VEVs of the $\phi_\alpha$}

In the case of the doublets $\phi_\alpha$,
the F terms from the superpotential~(\ref{W1})
only provide terms quadratic in the scalar components.
This is not problematic,
though,
because there are in this case quartic D terms
\be
V_D = 
\frac{g^2}{8}\, \sum_{i=1}^3 
\left[ \sum_\alpha \left( \phi_\alpha^\dagger \tau_i \phi_\alpha \right)
+ \phi_\nu^\dagger \tau_i \phi_\nu \right]^2 +
\frac{{g^\prime}^2}{8}
\left[ \sum_\alpha \left( \phi_\alpha^\dagger \phi_\alpha \right)
- \phi_\nu^\dagger \phi_\nu \right]^2,
\ee
(the $\tau_i$ are the Pauli matrices) which yield
\be
\langle 0 \left| V_D \right| 0 \rangle =
\frac{g^2 + {g^\prime}^2}{8}
\left( \left| v_e \right|^2 + \left| v_\mu \right|^2
+ \left| v_\tau \right|^2 - \left| v \right|^2 \right)^2.
\ee
Given the freedom in the
soft SUSY-breaking parameters,
which are subject only to the constraint
of breaking $S_3$ to the $\mu$--$\tau$ interchange symmetry,
there is enough freedom to adjust the VEVs $v_e$ and $v$.
It is sufficient for our purposes
to consider the VEVs $v_\mu$ and $v_\tau$.
One has to minimize a function with the structure
\be
\label{Fv}
G \left( v_{\mu, \tau} \right) =
2\, \mathrm{Re} \left[ f \left( v_\mu + v_\tau \right) \right]
+ b \left( \left| v_\mu \right|^2 + \left| v_\tau \right|^2 \right)
+ 2\, c\, \mathrm{Re} \left( v_\mu^\ast v_\tau \right)
+ d \left( \left| v_\mu \right|^2 + \left| v_\tau \right|^2 \right)^2,
\ee
where $d = \left( g^2 + {g^\prime}^2 \right) / 8$
as seen before.
The term $f \left( v_\mu + v_\tau \right)$ comes from
soft SUSY-breaking terms
of the form $\phi_{\nu,e}^\dagger \left( \phi_\mu + \phi_\tau
\right)$; $f$ is in general complex.

We want to show that it is possible to obtain
VEVs $v_\mu$ and $v_\tau$ \emph{different} from each other;
this is somehow the opposite of what was needed
for the $\chi_\alpha$.
Without loss of generality we use the parameterization
\be
v_\mu = u\, \cos{\sigma}, \quad 
v_\tau = \left( u\, \sin{\sigma} \right) e^{i \theta}
\quad ( u > 0, \ 0 < \sigma < \pi/2, \ \theta \in \mathbbm{R})
\ee
and obtain
\be
G =
2\, u \left[ \cos{\sigma}\, \mbox{Re}\, f
+ \sin{\sigma} \left| f \right| \cos{\left( \theta + \arg f \right)}
\right]
+ b u^2
+ 2 c u^2 \cos{\sigma} \sin{\sigma} \cos{\theta}
+ d u^4.
\ee
Since we only have to demonstrate that
$v_\mu \neq v_\tau$ is \emph{possible},
we are allowed to use some simplifying assumptions.
For $f$ real and positive and $c > 0$,
minimizing $G$
with respect to $\theta$
yields $\theta = \pi$.
Then one has
\be
\label{v-min}
\left. \frac{\partial G}
{\partial \sigma} \right|_{\theta = \pi}
= - 2 u \left( \cos{\sigma} + \sin{\sigma} \right) 
\left[ f
+ c u \left( \cos{\sigma} - \sin{\sigma} \right) \right] = 0.
\ee
We see that $v_\mu \neq v_\tau$ provided $f \neq 0$:
\be
\sin{\sigma} - \cos{\sigma} = \frac{f}{c u}.
\ee
In order to achieve $\left| v_\mu \right| \ll \left| v_\tau \right|$,
i.e.~$\cos{\sigma} \ll \sin{\sigma}$,
a finetuning is necessary:
$f \simeq c u$.

\section{Conclusions} \label{conclusions}

In this paper we have constructed a model
which realizes the Harrison--Perkins--Scott lepton mixing matrix. 
Our model is a supersymmetric extension
of an extended Standard Model,
based on the internal symmetry group $S_3$
and on seven $\mathbbm{Z}_2$ symmetries,
out of which six do not commute with $S_3$;\footnote{In other words,
the actual symmetry group is generated by the $S_3$
and six $\mathbbm{Z}_2$ symmetries,
and is thus much larger than $S_3$.}
three of these $\mathbbm{Z}_2$ symmetries
have a natural interpretation as discrete family lepton numbers.
The group $S_3$ simply permutes the families
$\alpha = e,\mu,\tau$ and easily allows to
describe the particle content of our model.
We have the left-handed
lepton doublets $\left( D_{L\alpha} \right)$
and the right-handed
charged-lepton singlets $\left( \alpha_R \right)$,
just as in the Standard Model;
furthermore,
there are four right-handed neutrino singlets
$\left( \nu_{R\alpha}, N_R \right)$,
four Higgs doublets $\left( \phi_\alpha, \phi_\nu \right)$,
and four scalar singlets with hypercharge zero
$\left( \chi_\alpha, S \right)$; 
the fields $N_R$,
$\phi_\nu$,
and $S$ are not affected by $S_3$.
Superpartners have to be supplied to all the fields above.

The following mechanisms are of importance for our model:
\begin{itemize}
\item For the generation
of each of the three charged-lepton masses $m_\alpha$
we need a separate Higgs doublet;
the different values of the three $m_\alpha$
are produced by the different VEVs $v_\alpha$ 
of the Higgs doublets.
In this way,
the charged-lepton mass matrix is automatically diagonal. 
\item The seesaw mechanism is responsible
for the smallness of the neutrino masses.
\item Lepton mixing originates in the mass matrix $M_R$
of the right-handed neutrino singlets.
In order to obtain the desired structure of $M_R$,
we need the $S_3$ invariance,
soft breaking of the discrete lepton numbers
in the $\nu_R$ Majorana mass terms,
and the special relation~(\ref{VEVs}) among the VEVs $w_\alpha$
of the scalar singlets $\chi_\alpha$. 
\item SUSY and the $\mathbbm{Z}_2$ symmetry $\mathbf{z}_\chi$
of~(\ref{zchi}) forbid terms
in the scalar potential which would destroy the relation~(\ref{VEVs}); 
SUSY is crucial
for an adequate separation of the Higgs-doublet terms 
from the scalar-singlet terms in the scalar potential,
such that we achieve three \emph{different} VEVs $v_\alpha$ while
the $w_\alpha$ fulfill the relation~(\ref{VEVs}).
\item A further mechanism to obtain the desired VEVs
is the soft breaking of $S_3$ down to $S_2$,
the $\mu$--$\tau$ interchange symmetry,
by terms of dimension 2.
\end{itemize}
We have taken care to demonstrate
that the VEV configuration required to obtain HPS mixing
corresponds to a local minimum of the scalar potential.

In our model the neutrino masses and the Majorana phases are \emph{free},
in contrast to lepton mixing which is completely fixed by construction.
In this sense,
the present model realizes a decoupling
of the mixing problem from the mass problem;
the latter is beyond the scope of the present model.

Since the charged-lepton mass matrix is diagonal,
flavour-changing neutral Yukawa interactions are absent 
\emph{at tree level} in the charged-lepton sector.
However,
such interactions appear at the one-loop level~\cite{loop}.
The specific nature of the lepton-flavour violation
makes $\mu^\pm \to e^\pm \gamma$ unmeasurably small,
but $\mu^\pm \to e^\pm e^+ e^-$ 
is possibly in the range of future experiments~\cite{loop}.

The symmetries invoked to enforce the HPS mixing scheme
hold at the seesaw scale.
Thus,
the renormalization group evolution down to the weak scale
will introduce corrections to the HPS mixing angles.
However,
it is well known that such corrections can only be sizeable
for a quasi-degenerate neutrino mass spectrum~\cite{balaji,xing}.
This was confirmed in~\cite{joshipura}
by the consideration of general perturbations
of a $\mu$--$\tau$ symmetric neutrino mass matrix.
In any case,
a first test of the HPS scheme
will be performed with the near-future experiments planned for
measuring $\left| U_{e3} \right|$---for a review and assessment of
these experiments see~\cite{huber}.

The model presented here uses an intricate interplay
of symmetries and mechanisms; a good part of this expenditure is used
for stabilizing the desired VEVs.
Our model is not,
though,
unduly
complicated when compared to other models
which produce HPS mixing.
This lets us wonder
whether HPS mixing is actually realized in Nature through symmetries
or it simply comes about by accident. 
Anyway,
the present model is the simplest one that we could construct
which realizes the basic idea
delineated in equations~(\ref{mdm})--(\ref{mmm})
of section~\ref{intro}.

\vspace*{5mm}

\paragraph{Acknowledgements:}
We thank Alfred Bartl,
Thomas Kernreiter,
and Jorge Crispim Rom\~ao for useful discussions on soft breaking and SUSY.
The work of L.L.\ was supported by the Portuguese
\textit{Funda\c c\~ao para a Ci\^encia e a Tecnologia}
through the projects POCTI/FNU/44409/2002
and U777--Plurianual.

\end{document}